\documentclass[12pt]{article}
\usepackage[pctex32]{graphics}
\usepackage[pctex32]{graphics}
\begin{document}
~
~
~
~~
\vspace{2cm}
\begin{center} {\Large \bf  Tachyon Tube on non BPS D-branes}
                                                  
\vspace{1cm}

                      Wung-Hong Huang\\
                       Department of Physics\\
                       National Cheng Kung University\\
                       Tainan, Taiwan\\

\end{center}
\vspace{1cm}
\begin{center} {\large \bf  Abstract} \end{center}
We report our searches for a single tubular tachyonic solution of regular profile on unstable non BPS D3-branes.  We first show that some extended Dirac-Born-Infeld tachyon actions in which new contributions are added to avoid the Derrick's no-go theorem still could not have a single regular tube solution.  Next we use the Minahan-Zwiebach tachyon action to find the regular tube solutions with circular or elliptic cross section.   With a critical electric field, the energy of the tube comes entirely from the D0 and strings, while the energy associated to the tubular D2-brane tension is vanishing.  We also show that fluctuation spectrum around the tube solution does not contain tachyonic mode.  The results are consistent with the identification of the tubular configuration as a BPS D2-brane.
\vspace{2cm}
\begin{flushleft}
E-mail:  whhwung@mail.ncku.edu.tw\\
\end{flushleft}
\newpage
\section {Introduction}

In an interesting paper [1], Mateos and Townsend showed that a cylindrical, or tubular D2-brane in a  Minkowski vacuum spacetime can be supported against collapse by the angular momentum generated by the Born-Infeld (BI) electric and magnetic fields.  Specifically, the Lagrangian density for unit surface tension is [1]
 $$ {\cal L} = -\sqrt{-\det(g+F)} .\eqno{(1.1)}$$
where $g$ is the induced metric on the 3D worldvolume and $F$ is the
worldvolume BI field strength 2-form.   The BI magnetic field $B$ is a worldspace scalar and the BI electric field  $E$ is chosen to be parallel  to the axis of the cylinder. Thus, the BI 2-form is
$$F= E\, dt\wedge dz + B\,  dz\wedge d\sigma\, .\eqno{(1.2)}$$
If we consider the cylindrical solution with radius $R$  the Lagrangian density becomes 
$$ {\cal L}(E,B,R) = -\sqrt{R^2(1-E^2) + B^2}\, .\eqno{(1.3)}$$
Introducing the electric displacement $ {\Pi} \equiv {\partial {\cal L}/ \partial E}$ the Hamiltonian density ${\cal H} \equiv {\Pi}E-{\cal L}$ becomes
$${\cal H}(\Pi,B,R) = R^{-1}\sqrt{\left(\Pi^2 + R^2\right)\left(B^2 + R^2\right)}\, .\eqno{(1.4)}$$
From this formula we see that the energy is minimized for given $B$ and $\Pi$
when  
 $$R= |B\Pi|.\eqno{(1.5)}$$
This is therefore the equilibrium value of the cylinder radius. The equilibrium
energy is
$${\cal H}_{min} = |\Pi| + |B|\, .\eqno{(1.6)}$$
This energy formula is typical of 1/4 supersymmetric configurations, and a
calculation confirms that the D2-brane configuration just describes preserves 1/4
supersymmetry, hence the name `supertube'.  More investigations about the supertube had been presented in [2].  The matrix theory interpretation was provided by Bak and Lee in [3] and others [4]. 

      In [5]  Kim et. al. had  investigated the supertube from the Dirac-Born-Infeld tachyon action, in the spirit of the Sen's conjecture that  the BPS branes can be viewed as tachyon kinks on non-BPS branes on higher dimension [6,7] - the remarkable `Decent Relation'.   The nontrivial coaxial array of tubular solution they found is the bound state of fundamental strings, D0-branes, and a cylindrical D2-brane and exhibit BPS-like property.

   In this paper we will report our searches for a single tubular tachyonic solution of regular profile on unstable non BPS D3-brane.  In section II we first briefly review the coaxial array of tubular solution [5] in the standard Dirac-Born-Infeld tachyon action. Then we turn to  investigate some extended models [8,9] in which new contributions are added to the tachyon action to avoid the Derrick's no-go theorem [10].  We show that, however, these extended models still do not provide us with a single regular tube solution.  In section III, we use the Minahan-Zwiebach tachyon action [11] to investigate the problem.   We have found a regular tube solution with circular cross section and seen that the energy of the single tubular configuration comes entirely from the D0 and strings which are on the D3-brane.  We calculate the fluctuation spectrum around the kink solution and see that there is no tachyonic mode therein.  In section IV, we present the similar calculations for the tube with elliptic cross section.   These results are consistent with the identification of the tubular configuration as a BPS D2-brane.      We make a conclusion  in the last section.

   Note that  the Minahan-Zwiebach tachyon action was successfully used by Hashimoto and Nagaoka [12] to show the phenomena of {\it kink condensation} and {\it vortex  condensation} in the unstable non BPS branes.   It supports the Sen's conjecture of the `Brane Descent Relations' of tachyon condensation.  Our investigations prove that the `Descent Relations' also show in the {\it tube condensation} on the unstable non BPS D3-brane. 

\section {Tubular Solutions in DBI Tachyon Action}
\subsection {Tube in Standard DBI Tachyon Action}
The standard Dirac-Born-Infeld (DBI) tachyon action for unstable D3-brane is [6]
$$ S = -{\cal T}_3 \int d^4x\; V(T) \sqrt{-\det (g_{\mu\nu} +
\partial_\mu T\partial_\nu T + F_{\mu\nu})} . \eqno{(2.1)}$$
In the cylindrical coordinate $ ds^2 = -dt^2 + dz^2 + dr^2 + r^2 d\theta^2$ and  the tachyon field we considered depends only on the radial coordinate $r$.   The BI electric and magnetic fields are taken to be 
$$ E_z = E,~~~F_{\theta z} = B,\eqno{(2.2)}$$
with constant $E$ and $B$.  Other components of EM field strength are vanishing.  In this case, the tachyonic effective action simplifies drastically
$$S =  -{\cal T}_3\int dt dz d\theta \int dr V(T) \sqrt{(1-E^2)r^2 + B^2}\sqrt{1+T'^2}\, ,\eqno{(2.3)}$$
where the prime ${}'$ denotes differentiation with respect to the radial
coordinate $r$. 

    To consider the supertube-like solutions we shall take $E=1$ [1] and thus work with the action
$$S =  - B {\cal T}_3\int dt dz d\theta \int dr \,V(T) \sqrt{1+T'^2} .\eqno{(2.4)}$$
The effective action above maps to that of a simple mechanical system with conserved ``energy'' by imagining $r$ as ``time'', which immediately  gives us the following integral of motion, 
$$ \frac{V(T)}{\sqrt{1+T'^2}} = C B/{\cal T}_3 , \eqno{(2.5)}$$
in which $C $ is an arbitrary integration constant.  Above equation may be rewritten as
$$ \frac{1}{2}T'^{2} + U(T)  = - {1\over 2} ,\eqno{(2.6)} $$
with 
$$ U(T) = - \frac{1}{2} \left[\frac{{\cal T}_3 V(T)}{C B}\right]^2 .\eqno{(2.7)} $$
Since the tachyon potential $V(T)$ measures varying tension, it shall satisfy two boundary values such that $V(T=0)=1$ and $V(T=\infty) = 0$ [6].  A nontrivial solution exists for ${{\cal T}_3/ C B}^{2} >1$.    In this case we have a coaxial  array of tubular configuration where $T$ oscillates as function of $r$.  The typical behavior of function U(T) is plotted in figure 1.
\\
\\
\scalebox{1}{\hspace{4cm}\includegraphics{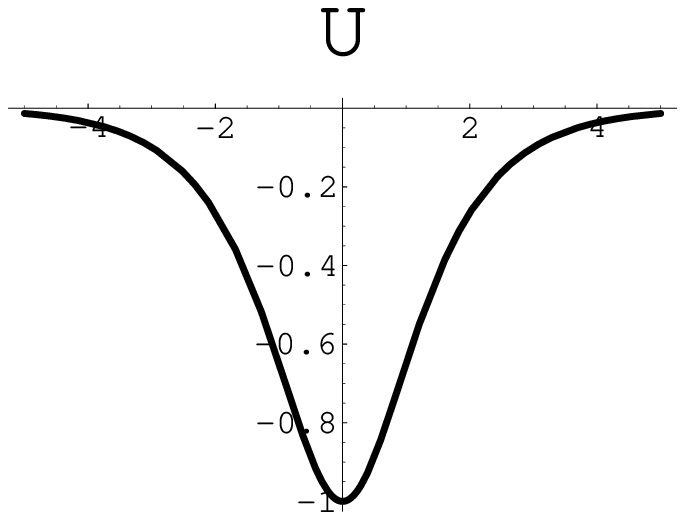}}\\ 
\hspace{2cm}{\it ~~~Figure.1. The general behaviors of  $U(T)$ in (2.7), (2.15), and (2.19). $U(T)$ is bounded between $0$ and $U^{min}$.  When $U^{min} < -1/2$ we could have a nontrivial solution in which $T$ oscillates as function of\, $r$. This corresponds to a coaxial array of tubular configuration.} \\
\hspace{1cm}

  For the specific form of tachyon potential [13]
$$V(T)=\frac{1}{\cosh \left(T/T_{0}\right)},\eqno{(2.8)}$$
where $T_{0}$ is $\sqrt{2}$ for the non-BPS D-brane in the superstring
and 2 for the bosonic string in the string unit, the exact solution $T(r)$ was found in [5].  After calculation the authors of [5] had shown that the energy of the tube comes entirely from the D0 and strings which are bounded on the D3-brane, as the relation (1.6) is shown.   This supports the Sen's conjecture of the `Brane Descent Relations' of tachyon condensation in the case of a coaxial  array of tubular configuration.

\subsection {Tube in Extended DBI Tachyon Action (I)}
     In this paper we attempt to investigate the problem for a single tube configuration. Following the literatures [8,9] we will add new contributions to the tachyon action.  Let us first consider the following extended DBI tachyon action
$$S =  - {\cal T}_3\int dt dz d\theta \int dr V(T) \left( \sqrt{(1-E^2)r^2+B^2}\sqrt{1+T'^2} + \epsilon T'^{2n} \right).\eqno{(2.9)}$$
The above action becomes the standard one (2.3) if the free parameter $\epsilon$ is vanishing.  The electric displacement and the energy density are 
$$\Pi =  {\cal T}_3 V(T) {r^2 E \sqrt{1+T'^2} \over \sqrt{(1-E^2)r^2+B^2}},\hspace{2cm}\eqno{(2.10)}$$
$$H  =  {\cal T}_3 V(T) \left[{\sqrt{r^2 + B^2} \sqrt{1+T'^2} \over \sqrt{(1-E^2)r^2+B^2}} + \epsilon T'^{2n} \right].\eqno{(2.11)}$$
Following Derrick's theorem [10], now we assume that there exists a kink profile $T(r)$ which extremizes the energy and consider the configuration obtained by dilating the coordinates, $r\rightarrow \lambda r$.  If the solution is to be an extremum of the energy then $({\partial \over \partial\lambda}\int H_\lambda)_{\lambda=1}=0$.   For the case of supertube, $E=1$, we then have the relation 

$$ {\cal T}_3\int dz d\theta dr V(T) \left[{2r^2\over B^2}\sqrt{1+T'^2} +{r^2+B^2 \over B^2 \sqrt{1+T'^2}} - (2n-1) \epsilon T'^{2n}\right]= 0.$$
\\
In the standard BDI action, $\epsilon =0$, then this relation forces $V=0$ or $T'^2 \rightarrow\infty$. This is the reasoning behind the a singular tube solution found in [14].  The singular solution complicates the analysis of its properties and it would therefore be useful to describe this solution as a limit of a regular solution.  Therefore we will consider the regularized action with $(2n-1) \epsilon > 0$.

    The Lagrangian in the case of supertube, $E=1$, is 
$$L  =  - B {\cal T}_3 V(T) \left(\sqrt{1+T'^2} + \epsilon T'^{2n}\right) .\eqno{(2.12)}$$
As before, the Lagrangian maps to that of a simple mechanical system with conserved ``energy'' by imagining $r$ as ``time'', which immediately  gives us the following integral of motion, 
$$ {\cal T}_3 V(T) \left[{B \over \sqrt{1+T'^2}}+ (1-2n) \,\epsilon \,T'^{2n}\right] = C  , \eqno{(2.13)}$$
in which $C $ is an arbitrary integration constant.  For the cases of $n=0 or 1$ the above equation may be rewritten as
$$ \frac{1}{2}T'^{2} + U(T)_n  = - {1\over 2} ,\eqno{(2.14)} $$
with 
$$ U(T)_{n=0} = - \frac{1}{2} {B^2\over \left[{C\over {\cal T}_3 V(T)} -\epsilon \right]^2} .\eqno{(2.15a)} $$
$$ U(T)_{n=1} = - \frac{1}{2} {B^2\over \left[{C\over {\cal T}_3 V(T)} \right]^2} - \epsilon B^2 {1 - B^2\left[{C\over {\cal T}_3 V(T)}\right]^2\over \left[{C\over {\cal T}_3 V(T)}\right]^4}+ O(\epsilon^2).\eqno{(2.15b)}$$
\\
As the tachyon potential $V(T)$ measures varying tension, it shall satisfy two boundary values such that $V(T=0)=1$ and $V(T=\infty) = 0$ [6].  This implies that 

$$ U(T)_{n=0} \rightarrow 0, ~~~~ as ~T\rightarrow \pm \infty,\eqno{(2.16a)}$$
$$ U(T)_{n=1} \rightarrow 0, ~~~~ as ~T\rightarrow \pm \infty.\eqno{(2.16b)}$$
\\
Therefore, denoting the minima value of  $U(T)_n$ as $U(T)_n^{min}$, then for the case of $U(T)_n^{min}  > -1/2$ the real solution $T$ does not exist. If  $U(T)_n^{min}  =  -1/2$  the real solution $T$ becomes a constant.  On the other hand, for the case of $U(T)_n^{min}  < -1/2$ the solution $T$ oscillates as function of $r$.    In this case we have a coaxial  array of tubular configuration (see the figure 1).  Thus we have proved that there does not exist a single, regular tube solution in the action (2.9) for the cases of $n=0 or 1$.  The cases with other values of $n$ remain to be analyzed. 

\subsection {Tube in Extended DBI Tachyon Action (II)}

   Let us next consider the following extended DBI tachyon action [8,9]
$$S =  - {\cal T}_3\int dt dz d\theta \int dr V(T) \left[(1-E^2)\,r^2+B^2\right]^{1/2 + \epsilon}\left[1+T'^2\right]^{1/2 + \epsilon}. \eqno{(2.17)}$$
The above action becomes the standard one (2.3) if the free parameter $\epsilon$ is vanishing. After performing the same analyzing as before we have the equation
$$ \frac{1}{2}T'^{2} + U(T)_\epsilon  = - {1\over 2} ,\eqno{(2.18)} $$
with 
$$ U(T)_\epsilon = - \frac{1}{2} \left(\frac{{\cal T}_3 V(T)}{C B}\right)^2 \left[1+ \epsilon \left[4 (lnB +1) - 4\left(\frac{{\cal T}_3 V(T)}{C B}\right)^2 + 2~ ln\left(\frac{{\cal T}_3 V(T)}{C B}\right)^2 \right]+ O(\epsilon^2) \right ] .\eqno{(2.19)} $$
in which $C $ is an arbitrary integration constant.  Therefore, as before, 
$$ U(T)_\epsilon \rightarrow 0, ~~~~ as ~T\rightarrow \pm \infty,\eqno{(2.20)}$$
and only if  $U(T)_\epsilon^{min}  < -1/2$ could we have a nontrivial solution.   However, the solution $T$ will oscillate as function of $r$ and we have a coaxial  array of tubular configuration (see the figure 1).  Thus we have also proved that the extended action (2.17) does not have a single, regular tube solution.

\subsection {Straight Kink and Tubular Kink on non BPS D-branes}
We conclude this section with a summary about the properties of  straight kink and tubular kink on non BPS D-branes.  

   (I)  The standard DBI tachyon can provide us with a singular straight kink [6,8] and a periodic array of straight, regular kink-antikink [15].  The extended DBI tachyon action with extra term $\epsilon T'^{2n}$ could provide us with a single, regular straight kink which is described by the tachyon field [8] 
$$T(x)  = {x\over (2n-1)\,\epsilon} ,\eqno{(2.21)}$$ 
The extended DBI tachyon action in which the Lagrangian modified as $\left(1+ T'^2\right)^{1/2 +\epsilon}$ could also provide us with a single, regular straight kink which is described by the tachyon field [8] 
$$T(x)  = {x\over \sqrt\epsilon}~ .\eqno{(2.22)}$$ 

   (II) The standard DBI tachyon action with Born-Infeld (BI) electric and magnetic fields can provide us with a singular tubular kink [14] and a periodic array of regular tachyon tubes [5].  However, neither the extended DBI tachyon action with extra term $\epsilon T'^{2n}$ nor the extended DBI tachyon action in which the power in the Lagrangian is modified as ${1/2 +\epsilon}$ could provide us with a single tube with regular profile, as was shown in this section.  

   As the singular solution complicates the analysis of its properties we will in the next section adopt the Minahan-Zwiebach tachyon action [11] to investigate the problem.   We can in there find a regular tube solution.

\section {Tube Solution in Minahan-Zwiebach Tachyon Action}

The Minahan-Zwiebach (MZ) tachyon action [11] is the low energy effective action which embodies the tachyon dynamics for unstable D-branes in (super)string theories.   Although the models were first proposed as a toy model capturing desirable properties of string theories, it was seen that the model is a derivative truncation of the BSFT action of the  non-BPS branes [16].    The MZ tachyon action was successfully used by Hashimoto and Nagaoka [12] to show the phenomena of {\it kink condensation} and {\it vortex  condensation} in the unstable non BPS branes.   It supports the Sen's conjecture of the `Brane Descent Relations' of tachyon condensation.  We have also used the MZ tachyon action to investigate the problems of the interaction between the kink-anti-kink configurations and recombination of intersecting branes [17].    

   The straight kink condensation in MZ tachyon action is easily to be seen.   As the Lagrangian is proportional to $V(T)\left(1+\left(\partial_x T\right)^2\right)$ the  equation of motion is  $V(T)' (1-T'(x)^2) = 2V(T) T''(x)$ which can be solved by a linear tachyon profile 
$$T = x. \eqno{(3.1)}$$
The solution interpolates two vacua of the theory at $T=\pm\infty$. The center of the kink sits at $x=0$.  After the analysis it is seen that there is no tachyonic fluctuation and the mass tower starts from a massless state and has the equal spacing [12].  It is a consequence that while the Derrick's theorem prevent us from obtaining a regular kink solution in the standard DBI tachyon action it does not forbid a regular kink solution in MZ tachyon action.  It is worthy to mention that the solution of tachyon profile $T=x$ is irrelevant to  the function form of the tachyon potential $V(T)$.  

\subsection {Tubular Kink in MZ Tachyon Action}

    The Minahan-Zwiebach tachyon action we use is [11,12]
$$ S = -{\cal T}_3 \int d^4x\; V(T) \left(1+ (\partial_\mu T)^2 + {1\over 4} F_{\mu\nu}^2\right) . \eqno{(3.2)}$$
In the cylindrical coordinate the above action with the BI electric and magnetic fields (2.2) becomes
$$ S = - 2\pi {\cal T}_3 \int dt\,dz\,dr\, r \,V(T) \left(1-{E^2\over2}+ {B^2\over 2r^2}+T'^2 \right) . \eqno{(3.3)}$$
The associated equation of motion is 
$$2V(T) \left(T''(x) + {T'\over r}\right) - V'(T)\left(1-{E^2\over2}+ {B^2\over 2r^2} - T'^2 \right) = 0, \eqno{(3.4)}$$
which can be solved by 
$$T(r)_c = {~B\over {\sqrt 2}}\, ln(r/r_0), \eqno{(3.5)}$$
if electric $E=E_c \equiv \sqrt 2$.  The value of  $r_0$ in above is an arbitrary integration constant.   Note that the value of $|T(r)_c|$ becomes zero at $r=r_0$ and the radius of tube will depend on the value of  $r_0$.   We will later see that the tube radius and $r_0$ is determined by the BI EM field, i.e., the charges of D0 and strings on the brane.  It is noted that, as that in the case of straight kink solution, the above tube solution is irrelevant to the function form of the tachyon potential $V(T)$.  Also, near the radius $r_0$ then $r- r_0 \equiv \epsilon  \ll 1$ and the solution (3.5) become $T(r)_c  \approx B\,\epsilon /  \sqrt 2$ which describe a linear profile likes that in the straight tachyon kink. 

  It is important to mention the physical meaning of the critical value of electric field $E_c$.  The action (3.3) tell us that the electric field has the effect of reducing the brane tension, and increasing $E$ to its `critical' value
$E_c=\sqrt 2$ would reduce the tension to zero if the magnetic field were zero; because that ${\cal L} \sim  B^2$ when $E=E_c$ (Note that from (3.5) we have a relation ${T'_c}^2 \sim B^2$).   This implies that the tachyon tube has no energy associated to the tubular D2-brane tension; its energy comes entirely from the electric and magnetic fields, which can be interpreted as `dissolved' strings and D0-branes, respectively. The energy from the D2-brane tension has been canceled by the binding energy released as the strings and D0-branes are dissolved by the D2-brane.  The phenomena that the tubular D2-brane tension has been canceled was crucial to have a supersymmetric tube configuration [1,18]. 

    To proceed we define the electric displacement defined by $\Pi = \partial L/\partial E$ and thus from (3.3) 
$$\Pi =  {\cal T}_3 V(T) \,r\, E ,\eqno{(3.6)}$$
The associated energy density defined by $H = \Pi E - L$  becomes
$$H  =  {\cal T}_3 V(T) \,r \left(2 + {B^2\over 2 r^2} \right),\eqno{(3.7)}$$
when $E=E_c$.  In figure 2 we plot the typical behaviors of function $H(r)$ which shows that there is a peak at finite radius.
\\
\\
\scalebox{1}{\hspace{4cm}\includegraphics{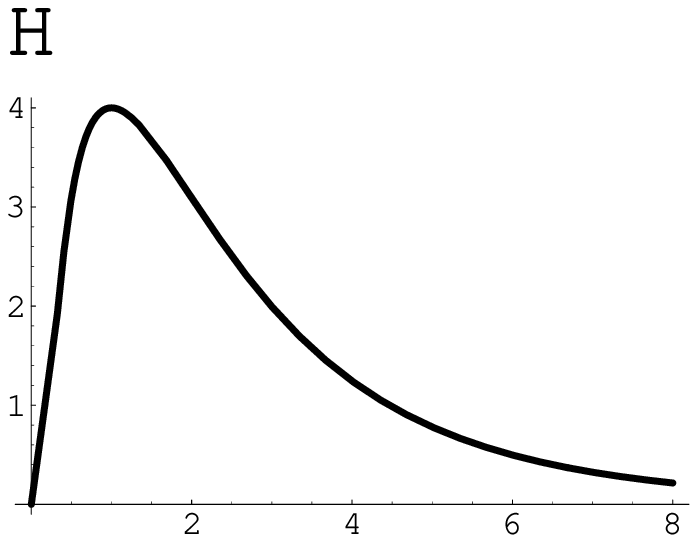}}\\ 
\hspace{2cm}{\it ~~~Figure.2. The typical behaviors of  $H(r)$ in (3.7).  There is a peak at finite radius r which specifies the size of the  circular cross-section tube.}\\
\hspace{1cm}

We can now us the regular tachyon solution (3.5) to evaluate the energy density ${\cal U}$ and string charge $q_1$.  The results are

$${\cal U} \equiv 2\pi \int_0^\infty dr H = 2\pi {\cal T}_3\left[\,{r_0^2\over {B\over T_0\sqrt 2}}\,\Gamma\left({2+{B\over T_0\sqrt 2}\over {B\over T_0\sqrt 2}}\right)\,\Gamma\left({{B\over T_0\sqrt 2}-2\over 2{B\over T_0\sqrt 2}}\right) + {B^2\over {2B\over T_0\sqrt 2}}  \right]. \hspace{1cm} \eqno{(3.8)}$$
\\
$$q_1 \equiv 2\pi  \int_0^\infty dr \,\Pi =2\pi {\cal T}_3\left[\,{r_0^2 \sqrt 2\over {2B\over T_0\sqrt 2}}\,\Gamma\left({2+{B\over T_0\sqrt 2}\over {B\over T_0\sqrt 2}}\right)\,\Gamma\left({{B\over T_0\sqrt 2}-2\over 2{B\over T_0\sqrt 2}}\right)\right]. \hspace{2.2cm}\eqno{(3.9)}$$
\\
in which the functional form of tachyon potential $V(T)$ in (2.8) is used.  We can now follow [5] to define the $D_0$ charge $q_0$ by the relation 
$$q_0 \sim  B\, T_0\,{\cal T}_3, \eqno{(3.10)}$$
then
$${\cal U} =\sqrt 2 \,q_1 +  \,q_0 , \eqno{(3.11)}$$
and when $B\gg T_0 \sqrt 2$ equation (3.9) implies that 
$$ q_1 \sim r_0^2\, \,q_0^{-1} \, \,\,\,\Rightarrow\,\,\,\,r_0 \sim \sqrt {\,q_1 \,q_0},\eqno{(3.12)}$$
which is like the relation (1.5).   Thus we have seen that the tubular D2-brane tension has been canceled by the binding energy released as the strings and D0-branes are dissolved by the D2-brane.   In this interesting case, i.e. $E=E_c$ the energy of the tube comes entirely from the D0 and strings which are bounded on the D3-brane. 

   We conclude this section a comment. The critical electric in DBI action is $E_c=1$ [1] while that in the MA tachyon action is $E_c=\sqrt 2$.  In the later case the tube profile is described in (3.5) and does not depend on the functional form of tachyon potential $V(T)$.  The inconsistence may be traced to the fact that the MZ tachyon action is just a derivative truncation of the BSFT action of the  non-BPS branes [16].   However, we hope that the truncated action could capture desirable properties of the brane theories.

\subsection{Fluctuation around Tubular Kink}
Let us now consider the fluctuation $t$ around the tubular solution 
$$T(r) = T(r)_c + t(r).  \eqno{(3.13)}$$
Substituting the tubular kink solution $T(r)_c$ in (3.5) into the action (3.3) and considering only the quadratic terms of fluctuation field $t(r)$ we obtain 
$$S = - 2\pi\,{\cal T}_3 \int dr\, r \left(V(T_c)\,t'^2 + 2V(T_c)'\,t\,t' +{V(T_c)''\over2}\left( {B^2\over 2r^2}+(T'_c)^2\right) t^2 \right) $$
$$ = - 2\pi\,{\cal T}_3 \int dr\,  \left( r\, \hat{t'}^2\, + \left( -{B^2\over 2r} + {B^4\over 4r} \,\,\left(ln\left({r/ r_0}\right) \right)^2\right) \hat{t}^2 \right) . \hspace{2.2cm}\eqno{(3.14)}$$
\\
in which we have used an partial integration and the field redefinition 
$$\hat{t} \equiv V(T_c)^{1\over2} \,t,~~~~~~~~~with ~~~~V(T_c) = e^{-T_c^2}. \eqno{(3.15)}$$
After defining a new variable 
$$w= ln\left({r/ r_0}\right),\eqno{(3.16)}$$
we obtain
$$S =- 2\pi\,{\cal T}_3 \int dw\, \left[ \partial_w\hat{t}\, \partial_w\hat{t} + \hat{t}\left( -{B^2\over 2} + {B^4\over 4} \, w^2  \right)\hat{t}\right]. \eqno{(3.17)}$$
From the above expression we see that the fluctuation $t$ obeys a Schr\"odinger equation of a harmonic oscillator, thus the mass squared for the fluctuation is equally spaced and specified by an integer $n$, 
$$m^2_t=  2n\,B^2 , ~~~~~ \ n \geq 0 .\eqno{(3.18)}$$
Thus there is no tachyonic fluctuation, the mass tower starts from a massless state and has the equal spacing. This result is consistent with the identification of the tachyon tube as a tubular BPS D2-brane. 

\section{Elliptical Tube in MZ Action}
Historically, following the initial supertube paper [1], a matrix model version of it was introduced by Bak and Lee [3]. A subsequent paper by Bak and Karch [4] found a more general solution of the matrix model describing an elliptic supertube, which included a plane parallel D2/anti-D2 pair as a limiting case.  The fact that a circular cross-section could be deformed to an ellipse was a surprise at first sight because it was hard to understand how any shape other than a circle could be consistent with both rotation and the time-independent energy profile required by supersymmetry.  The subtleties was detailed by Townsend in [18].   In this section we will apply the similar calculations as those in the circular cross-section to the tube with elliptic cross section.   We will see that deform a circular cross-section tachyon tube to that with an elliptical cross-section does not change the physical properties.

    The elliptical cylinder can be described by the coordinate ($u$,$v$,$z$), with $0 \leq u <\infty$, $0 \leq v < 2\pi$, and $- \infty < z < \infty$.  Define $x \equiv a \,cosh(u)\,cos(v)$,\,\,$y \equiv a \,sinh(u)\,sin(v)$ implies that $x^2/cosh(u)^2 + y^2/sinh(u)^2 = a^2$ and $x^2/cos(v)^2 - y^2/sin(u)^2 = a^2$. Thus holding $u$ constant yields a family of ellipses with x-axial the major one and holding $v$ constant yields a family of hyperbolas with focal on the x-axial.  The line element becomes
$$ds^2 = -dt^2 + a^2 \left[sinh(u)^2+ sin(v)^2\right] (du^2 +dv^2) + dz^2.\eqno{(4.1)}$$
Substituting the BI 2-form  
$$F= E\, dt\wedge dz + B\,  dz\wedge dv\, .\eqno{(4.2)}$$
into MZ tachyon action (3.2) we have the action
$$ S = -  {\cal T}_3 \int dt\,dz\,du\, dv \,V(T) \left(a^2\,\left(1-{E^2\over2}\right) \left[sinh(u)^2+ sin(v)^2\right]+ {B^2\over 2}+T'^2 \right) .$$
$$= - 2\pi {\cal T}_3 \int dt\,dz\,du \,V(T) \left(a^2\,\left(1-{E^2\over2}\right) \left[sinh(u)^2 + {1\over2}\right]+ {B^2\over 2}+T'^2 \right), \eqno{(4.3)}$$
in which $T' \equiv \partial_u T(u)$.  The associated equation of motion is 
$$2V(T) \, T''(u)  - V'(T)\left(a^2\,\left(1-{E^2\over2}\right) \left[sinh(u)^2 + {1\over2}\right] + {B^2\over 2} - T'^2 \right) = 0, \eqno{(4.4)}$$
which can be solved by 
$$T(u)_c = {~B\over {\sqrt 2}}\, (u - u_0), \eqno{(4.5)}$$
if electric $E=E_c \equiv \sqrt 2$. Note that  $u_0$ being determined by the BI electric and magnetic fields is relevant to the size of the ellipse.

  As that in the circular cross-section, we can from the action (4.3) see that increasing the electric field has the effect of reducing the brane tension, and increasing $E$ to its `critical' value $E_c=\sqrt 2$ would reduce the tension to zero if the magnetic field were zero.   The electric displacement and energy density defined as before become

$$\Pi = 2\pi {\cal T}_3  \,V(T) \,a^2\, E  \left[sinh(u)^2 + {1\over2}\right], \hspace{4.1cm}\eqno{(4.6)}$$
$$H  = 2\pi {\cal T}_3 V(T) \left(a^2\,\left(1+{E^2\over2}\right) \left[sinh(u)^2 + {1\over2}\right]+ {B^2\over 2}+T'^2 \right). \eqno{(4.7)}$$
\\
In figure 3 we plot the typical behaviors of the function $H(u)$ which shows that there is a peak at finite radius.
\\
\\
\scalebox{1}{\hspace{4cm}\includegraphics{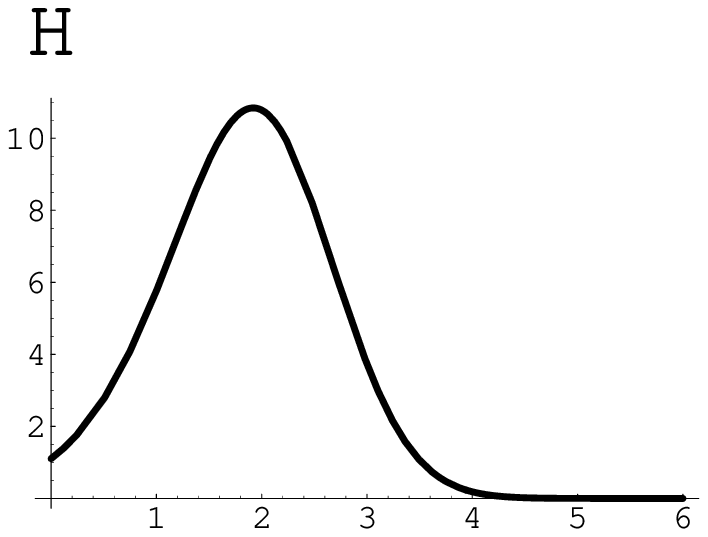}}\\ 
\hspace{2cm}{\it ~~~Figure 3. The typical behaviors of $H(u)$ in (4.7).  There is a peak at finite u which specifies the size of the elliptic tube.}\\
\hspace{1cm}

When $E=E_c$, we can use the regular tachyon solution (4.5) to evaluate the string charge per unit length of the elliptic tube
$$q_1 = 2\pi  \int_0^\infty du \,\Pi, \eqno{(4.8)}$$
which then determines the $u_0$ in terms of $B$ and $q_1$.  After the analysis we can see that the energy density per unit length ${\cal U}$ can be expressed as
$${\cal U} \equiv 2\pi \int_0^\infty du\, H = \sqrt 2 q_1 +  q_0 ,\eqno{(4.9)}$$
in which $q_0$ being the charge of $ D_0$ is the function of the magnetic field $B$ [5].  (Note that, in the elliptic case we could not perform the above integrations exactly if $u_0 \not =0$.)   The results tell us that, as that in the circular cross-section solution,  the tubular D2-brane tension has been canceled by the binding energy released as the strings and D0-branes are dissolved by the D2-brane.   In this interesting case, i.e. $E=E_c$ the energy of the tube comes entirely from the D0 and strings which are bounded on the D3-brane.  

    Let us now turn to investigate the fluctuation $t$ around the elliptical tubular solution 
$$T = T(u)_c + t(u).  \eqno{(4.10)}$$
Substituting the tubular solution (4.5) into the action (4.3) and considering only  the quadratic terms of fluctuation field $t(r)$ we obtain 
$$S = - 2\pi\,{\cal T}_3 \int du\, \left(V(T_c)\,t'^2 + 2V(T_c)'\,t\,t' +{V(T_c)''\over2}\left({B^2\over 2}+(T'_c)^2\right) t^2 \right) $$
$$ = - 2\pi\,{\cal T}_3 \int du\, \left[ \partial_u\hat{t}\, \partial_u\hat{t} + \hat{t}\left( -{B^2\over 2} + {B^4\over 4} \, u^2  \right)\hat{t}\right], \hspace{3cm}\eqno{(4.11)}$$
in which we have used an partial integration and the field redefinition 
$$\hat{t} \equiv V(T_c)^{1\over2} \,t,~~~~~~~~with~~~V(T_c) = e^{-T_c^2}. \eqno{(4.12)}$$
The above expression is exactly like (3.17) and have the same spectrum as (3.18).  Thus there is no tachyonic fluctuation. This result is consistent with the identification of the elliptic tachyon tube as a tube BPS D2-brane. 


   We conclude this section with a comment.   The elliptical tube described by the solution (4.5) does not depend on the function form of the tachyon potential. However, in the case of describing the non DPS D-brane by the tachyon potential (2.8), as there exist term $sinh(u)^2$ in (4.7) the tube could be with a finite energy density only if $B>2\sqrt2$.  Otherwise, after the integration the equation (4.9) will be divergent.   However, if we described the non DPS D-brane by the tachyon potential $V(T) = e^{-T^2}$, then there does not have such a constrain.

\section{Conclusion}

We have reported our searches for a single tubular solution of regular profile on unstable non BPS D3-brane.  We have shown that some extended models in which new contributions are added to the tachyon action to avoid the Derrick's no-go theorem still could not provide us with a single regular tube solution.  We thus use the Minahan-Zwiebach tachyon action to investigate the problem.  We have found regular tubes of circular and elliptic cross-section and seen that the energy of the single tubular configuration comes entirely from the D0 and strings which are on the D3-brane.  We have calculated the fluctuation spectrum around the kink solution and seen that there is no tachyonic mode therein. The results are consistent with the identification of the tubular configuration as a BPS D2-brane.      These provides us with a new support of the Sen's Decent Relations of tachyon condensation.

\newpage
 
\begin{center} {\large \bf  References} \end{center}
\begin{enumerate}
\item D. Mateos and P. K. Townsend, ``Supertubes'', Phys. Rev. Lett. 87 (2001) 011602 [hep-th/0103030]; R. Emparan, D. Mateos and P. K. Townsend, ``Supergravity Supertubes'', JHEP 0107 (2001) 011 [hep-th/0106012]; D.~Mateos, S.~Ng and P.~K.~Townsend, ``Tachyons, supertubes and brane/anti-brane systems'', JHEP  0203 (2002) 016 [hep-th/0112054].
\item  M. Kruczenski, R. C. Myers, A. W. Peet, and D. J. Winters,``Aspects of supertubes'',  JHEP 0205 (2002) 017 [hep-th/0204103];Y. Hyakutake and N. Ohta, ``Supertubes and Supercurves from M-Ribbons,'' Phys. Lett. B539  (2002) 153 [hep-th/0204161]; N. E. Grandi and A. R. Lugo, ``Supertubes and Special Holonomy'', Phys. Lett. B553 (2003) 293 [hep-th/0212159]; B. Cabrera Palmer and D. Marolf , `` Counting Supertubes'',  JHEP 0406 (2004) 028 [hep-th/0403025]; D. Bak, Y. Hyakutake, and N. Ohta, ``Phase Moduli Space of Supertubes,'' [hep-th/0404104];  Wung-Hong Huang, `` Condensation of Tubular D2-branes in Magnetic Field Background'' [hep-th/0405192].
\item D. Bak, K. M. Lee, ``Noncommutative Supersymmetric Tubes'',  Phys. Lett. B509 (2001) 168 [hep-th/0103148].
\item D. Bak and S. W. Kim, ``Junction of Supersymmetric Tubes,'' Nucl. Phys.  B622 (2002) 95 [hep-th/0108207]; D. Bak and A. Karch, ``Supersymmetric Brane-Antibrane Configurations,'' Nucl. Phys.  B626 (2002) 165 [hep-th/011039]; D. Bak and N. Ohta, ``Supersymmetric D2-anti-D2 String,'' Phys. Lett.  B527 (2002) 131 *[hep-th/0112034]; D. Bak, N. Ohta and M. M. Sheikh-Jabbari, ``Supersymmetric Brane-Antibrane Systems: Matrix Model Description, Stability and Decoupling Limits,'' JHEP  0209 (2002) 048 [hep-th/0205265].
\item C. Kim, Y. Kim, O-K. Kwon, and P. Yi, ``Tachyon Tube and Supertube,''  JHEP 0309 (2003) 042 [hep-th/0307184].
 \item A. Sen, ``Tachyon Condensation on the Brane Antibrane System'', JHEP 9808 (1998) 012, [hep-th/9805170]; ``Descent Relations Among Bosonic D-branes'',  Int.\ J.\ Mod.\ Phys. A14 (1999)  4061, [hep-th/9902105];  ``Non-BPS States and Branes in String Theory'',  [hep-th/9904207];  ``Universality of the Tachyon Potential'',  JHEP 9912 (1999) 027, [hep-th/9911116]; ``Supersymmetric world-volume action for non-BPS D-branes,''  JHEP  9910 (1999) 008 [hep-th/9909062].
\item A. Sen, ``Dirac-Born-Infeld Action on the Tachyon Kink and Vortex,''  Phys. Rev. D68 (2003) 066008 [hep-th/0303057].
\item E. J. Copeland, P. M. Saffin, and D. A. Steer, ``Singular tachyon kinks from regular profiles'', Phys. Rev. D68 (2003) 065013  [hep-th/0306294]. 
\item D. Bazeia, R. Menezes, and J.G. Ramos,  ``Regular and Periodic Tachyon Kinks [hep-th/0401195].
\item  G.  Derrick, ``Comments On Nonlinear Wave Equations As Models For         Elementary Particles'',  J.Math.Phys.  5 (1964) 1252. 
\item  J.  A. Minahan and B.  Zwiebach,``Gauge Fields and Fermions in Tachyon Effective Field Theories'',  JHEP 0102 (2001) 034 [hep-th/0011226]; ``Effective Tachyon Dynamics in Superstring Theory'',  JHEP  0103 (2001) 038 [hep-th/0009246].
\item K. Hashimoto and S. Nagaoka,``Realization of Brane Descent Relations 
in Effective Theories'', Phys. Rev. D66 (2002) 0206001  [hep-th/0202079].
\item  D. Kutasov and V. Niarchos, ``Tachyon effective actions in open string theory,''  [hep-th/0304045].
\item  L.~Martucci and P.~J.~Silva, ``Kinky D-branes and straight strings of open string tachyon effective  theory,''  JHEP 0308 (2003) 026 [hep-th/0306295]
\item C. Kim, ``Tachyon Kinks,''  JHEP 0305 (2003) 020 [hep-th/0304180]; P. Brax, J.Mourad, and D.A.Steer,  ``Tachyon kinks on non BPS D-branes'', Phys.Lett. B575 (2003) 115 [hep-th/0304197].    
\item  B.  Zwiebach, ``A Solvable Toy Model for Tachyon Condensation in String Field Thoery' , JHEP 0009 (2000) 028 [hep-th/0008227];  J.  A.  Minahan and B.  Zwiebach, ``Field Theory Models for Tachyon and Gauge Field String
Dynamics'',   JHEP 0009 (2000)  029 [hep-th/0008231]; O.   Andreev, ``Some Computations of Partition Functions and Tachyon Potentials in Background Independent Off-Shell String Theory'' ,  Nucl.Phys. B598 (2001) 151 [hep-th/0010218].
\item Wung-Hong Huang, ``Brane-Antibrane Systems Interaction under Tachyon Condensation'' , Phys.Lett. B561 (2003) 153 [hep-th/0211127]; Wung-Hong Huang, ``Recombination of Intersecting D-branes in Tachyon Field Theory'' , Phys.Lett. B564 (2003) 155 [hep-th/0304171]. 
\item P. K. Townsend, ``Surprises with Angular Momentum'', Annales Henri Poincare 4 (2003) S183 [hep-th/0211008].
\end{enumerate}
\end{document}